\documentclass[reprint,amsmath,amssymb,pra,twocolumn,longbibliography]{revtex4-1} 
\usepackage{graphicx}
\usepackage{color}
\usepackage{dcolumn}

\newcommand{\vare}{\varepsilon}
\newcommand{\abs}[1]{\left| #1 \right|}

\begin{document}

\preprint{APS/123-QED}

\title{Exceptional points of Bloch eigenmodes on a dielectric slab\\
  with a periodic array of cylinders} 

\author{Amgad Abdrabou}
\email{Corresponding author: mabdrabou2-c@my.cityu.edu.hk}

\affiliation{Department of Mathematics, City University of Hong Kong,
  Kowloon, Hong Kong, China}

\author{Ya Yan Lu}

\affiliation{Department of Mathematics, City University of Hong Kong,
  Kowloon, Hong Kong, China}

\date{\today}

\begin{abstract}
Eigenvalue problems for electromagnetic resonant states on open
dielectric structures are non-Hermitian and may have exceptional
points (EPs) at which two or more eigenfrequencies and the
corresponding eigenfunctions coalesce. EPs of resonant states for
photonic structures give rise to a number of unusual wave phenomena
and have potentially important 
applications. It is relatively easy to find a few EPs for a structure
with parameters, but isolated EPs provide no information about their formation and
variation in parameter space, and it is always difficult to ensure
that all EPs in a domain of the parameter space are found. In this
paper, we analyze EPs for a dielectric  slab containing a periodic
array of circular cylinders. By tuning the periodic structure towards a
uniform slab and following the EPs continuously, we are able to obtain
a precise condition about the limiting uniform slab, and thereby order
and classify EPs as tracks with their endpoints determined analytically. 
It is found that along each track, a second order EP of resonant
states (with a complex frequency) is transformed to a special kind of
third order EP with a real frequency via a special fourth order EP. 
Our study provides a clear and complete picture for EPs in parameter
space, and gives useful guidance to their practical applications.  
\end{abstract}

\maketitle


\section{Introduction}
\label{S1}

A non-Hermitian eigenvalue problem (EVP) often has multiple eigenvalues sharing
a single linearly independent eigenfunction. For EVPs 
depending on parameters, the set of parameter values corresponding to
such a non-Hermitian degeneracy  is called an exceptional point
(EP)\cite{Kato,Heiss,Berry}.  
In the photonics community, there is currently a significant research effort to 
construct systems with EPs, find their unusual properties, and realize
their applications \cite{Miri19,Ozdemir19}. The associated EVPs are either 
directly formulated from the governing Maxwell's equations or
indirectly formulated from the scattering operators.
For parity-time (${\cal PT}$) symmetric photonic systems \cite{Ozdemir19, BenderPT,Feng17}, 
EPs can be easily found by tuning a single parameter, namely, the
magnitude of the the balanced gain and loss. 
A number of interesting wave phenomena have been observed
in photonic systems with ${\cal PT}$ symmetry and EPs
\cite{PTsym,EPLas2}. Potential applications of EPs include
lasing \cite{EPLas3,Feng}, sensing
\cite{Wiersig,Hodaei,Chen2017,Langbein,Zhang19}, robust mode 
switching \cite{Doppler16,Hassan}, etc. 

Since a balanced gain and loss is not always desirable, it is of
significant interest to explore EPs in open photonic systems without material loss and
artificial gain. Due to the possible radiation losses,  EVPs for open systems are
non-Hermitian and can have EPs 
\cite{BoZhen,Arslan,Zhou2018,Kullig18,Amgad1,Amgad3}. 
For Maxwell's equations, the standard EVP formulation regards the
frequency $\omega$ as the eigenvalue. For open systems, the solutions
of the Maxwell EVP are resonant states (also called resonant modes or
quasi-normal modes) with complex frequencies 
\cite{Tsuji83,Glisson83,Fan,Powell14,Yan18,Amgad2}. 
%
%
A second order EP of resonant states corresponds to the simultaneous
coalescence of two complex eigenfrequencies and the corresponding
eigenfunctions, and typically requires the tuning of two real
parameters. Therefore, second order EPs are isolated points in
the plane of two real parameters. On bi-periodic structures such as
photonic crystal slabs, EPs 
are relatively easy to find, since the resonant states are Bloch waves
and the two components of the Bloch wavevector serve as parameters
\cite{BoZhen,Arslan}. For two-dimensional (2D) structures that are
invariant in one spatial direction and periodic in another, second
order EPs of resonant states (in $E$ or $H$ polarization) can be
found by tuning one structural parameter and the Bloch wavenumber
\cite{Amgad1}. 

Finding a few isolated EPs in a plane of two parameters is relatively
easy, but it is difficult to 
ensure that
all EPs in a given domain of
parameter space are determined. Moreover, for realizing potential applications,
it is highly desirable to 
understand how EPs depend on parameters of the structure.
One approach is to introduce an additional parameter that simplifies
the structure in a proper limit, follow the EPs to the limit, and
find and analyze the limiting structure. 
In Ref.~\cite{Amgad1}, we studied EPs of resonant states for a 
periodic slab with two rectangular segments in each period, followed
the EPs as the thickness of the one segment tends to zero, and
classifies EPs as connected sets in parameter space indexed by an 
integer pair $(m,n)$. 
However, when the width of the one segment is very small, 
it is difficult to obtain accurate numerical solutions, because the related linear
systems become ill-conditioned. Consequently, we were not able to
determine a precise condition for the limiting structure. 

In this paper, we consider a slab with a periodic array of circular 
cylinders, and follow the EPs as the refractive index of the cylinders
tends to that of the slab. A numerical method is proposed to study the
eigenmodes and EPs of the periodic structure, and it remains
well-conditioned even when the refractive indices of the cylinders
and the slab are nearly equal. The high accuracy numerical solutions
reveal an interesting transition from second order to third
order EPs via a fourth order EP, and allows us to find a 
precise condition for the limiting uniform slab.

The rest of this paper is organized as follows. In Sec.~\ref{S2}, we
compare the band structures of a uniform slab and a periodic slab,
discuss different kind of eigenmodes and the so-called intrinsic EPs. 
In Sec.~\ref{S3}, we present a track of EPs as a periodic slab
approaches its uniform limit, and show the different types of EPs on
the track. In Sec.~\ref{S4}, we analyze the limiting uniform slab, and
show additional tracks of EPs.  The paper is concluded with some remarks
in Sec.~\ref{S5}.

\section{Band structures of uniform and periodic slabs}
\label{S2}

\begin{figure}[htbp]
  \centering
\includegraphics[width=0.85\linewidth]{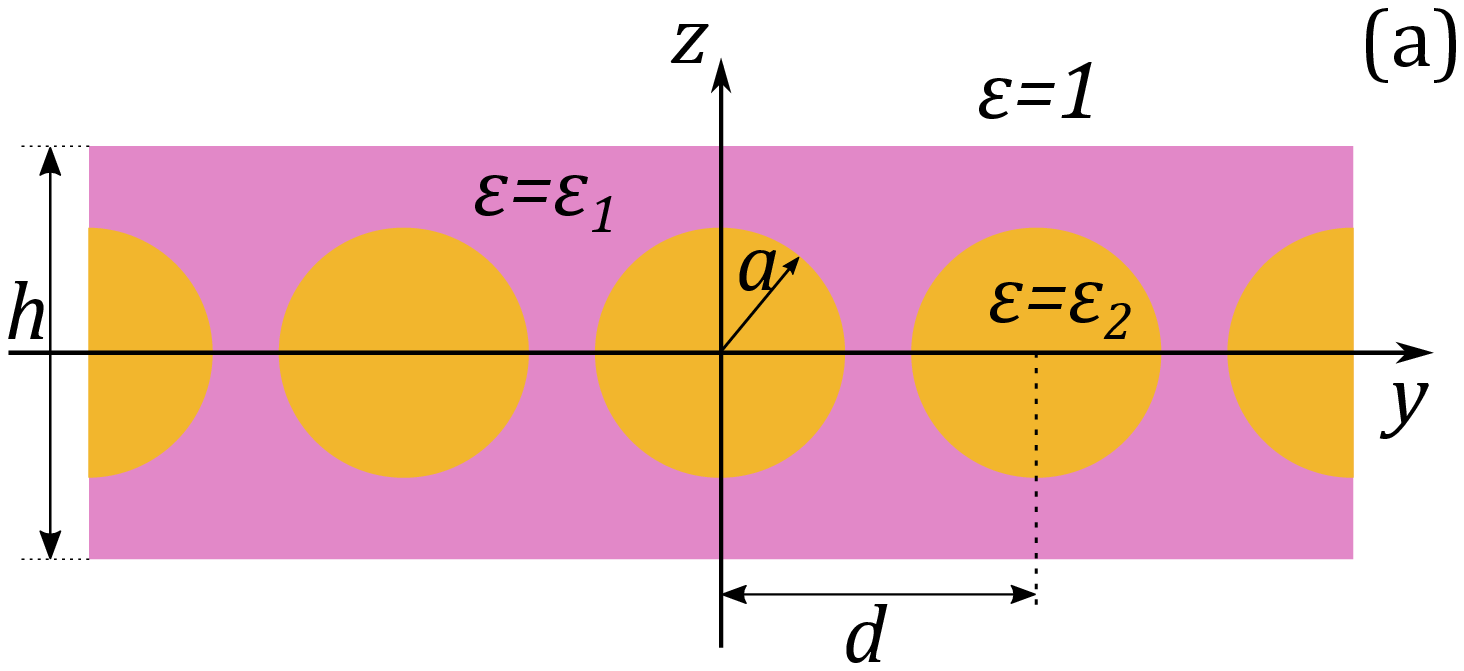}\\
\vspace{0.5cm} 
\includegraphics[width=0.85\linewidth]{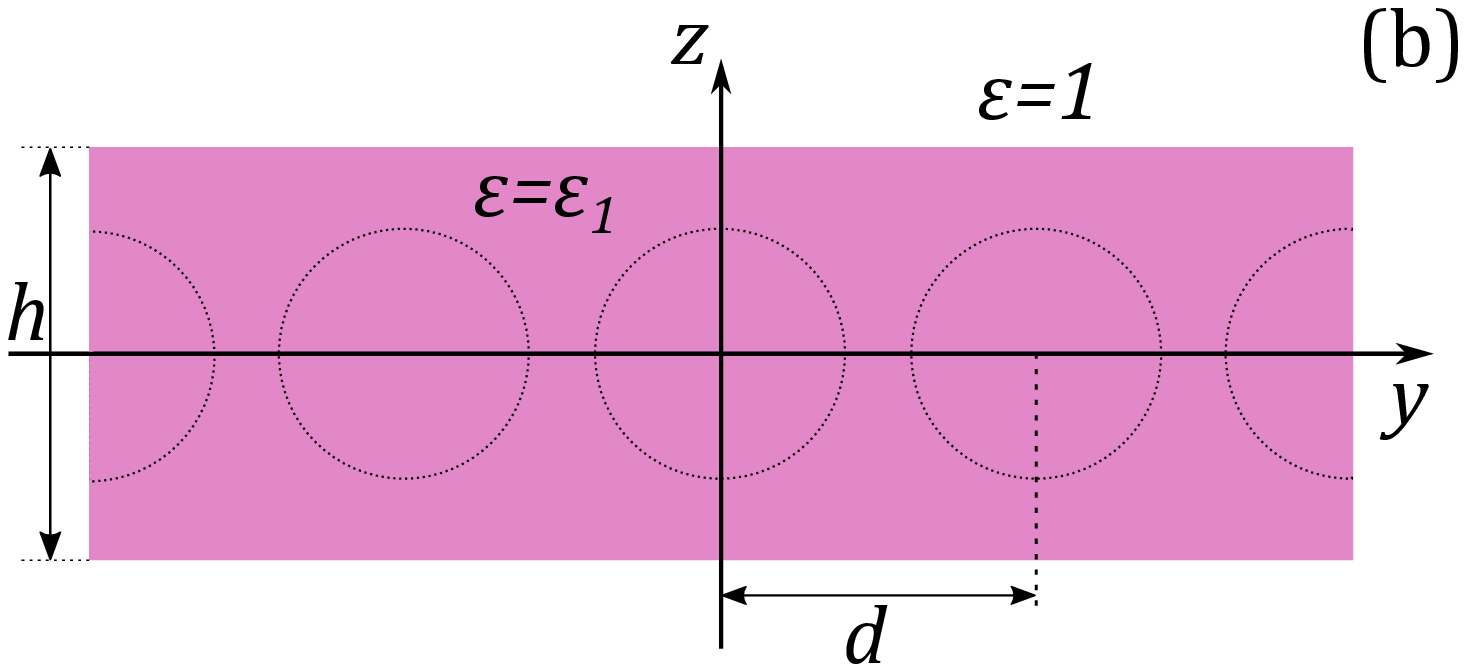} 
  \caption{(a) A dielectric slab with a periodic array of circular
    cylinders. (b) A uniform slab with a fictitious period.}
  \label{fig1ab}
\end{figure}

In Fig.~\ref{fig1ab}(a), 
we show a periodic  array of circular cylinders  with radius $a$ and dielectric constant $\varepsilon_2$ embedded in a slab  with dielectric constant $\varepsilon_1$. 
The structure has a period $d$ and a thickness $h$, and is surrounded  by vacuum. A Cartesian coordinate system is chosen such that the structure is invariant in $x$, periodic in $y$, and symmetric in $z$. The dielectric function
satisfies $\vare(y,z) = 1$ for $\abs{z}> h/2$. 
A uniform slab with  the same dielectric constant $\varepsilon_1$ and
thickness $h$  is shown in Fig.~\ref{fig1ab}(b).
We are interested in EPs of resonant states for this periodic slab, especially  when 
$\varepsilon_2 \to \varepsilon_1$. For the purpose of comparison,
the uniform slab is also regarded as a periodic structure with a
fictitious period $d$. 


For simplicity, we consider only the 
$E$-polarization for which the $x$-component of the electric field (denoted as $u$) satisfies the 
following 2D Helmholtz equation 
\begin{equation}
\label{fEq1}
	\partial_y^2 u +\partial_z^2 u +k^2 \vare(y,z)\, u = 0, 
\end{equation}
where $k=\omega/c$ is the freespace wavenumber, $\omega$ is the
angular frequency, $c$ is the speed of light in vacuum, and the time
dependence is $\exp(-i\omega t)$.
A Bloch mode on the periodic slab is a solution of Eq.~(\ref{fEq1}) given as
\begin{equation}
\label{fEq2}
	u(y,z) = \phi(y,z)\,e^{i\beta y},
\end{equation}
where $\phi(y,z)$ is periodic in $y$ with period $d$ and $\beta$ is
the Bloch wavenumber.  
If $u$ decays exponentially to zero as $z\to \pm \infty$, the Bloch
mode is a guided mode. 
If power is radiated to infinity, i.e., $u$ satisfies outgoing
radiation  conditions as $z\to \pm \infty$, then the Bloch mode is a
resonant state. We consider only dielectric structures with a real
$\vare$, i.e., no material loss or gain. In that case, a guided mode has a 
real $\omega$ and a real $\beta$, a resonant state has a real $\beta$
and a complex $\omega$ with a negative imaginary part.

Corresponding to
each Bloch mode $\{ u, k, \beta\}$, there is a
reciprocal mode at the 
same frequency for Bloch wavenumber $-\beta$ and a field distribution 
$v(y,z)$. Since the structure is symmetric in $y$, we can assume $v(y,z) =
u(-y,z)$.  If $\{ u, k,
\beta \}$ is a resonant state with a real $\beta$ and 
$\{ v, k, -\beta \}$ is its reciprocal mode, then the complex conjugate of
$v$, denoted as $\overline{v}$, satisfies the Helmholtz equation with
$k$ replaced by $\overline{k}$. In addition, $\overline{v}$ satisfies
an incoming wave condition at infinity (the opposite of the outgoing
radiation condition), and the associated Bloch wavenumber is $\beta$. As
in \cite{Amgad2}, we call this solution $\{ \overline{v}, \overline{k},
\beta\}$, the time reversal of the resonant state $\{ u, k, \beta
\}$.
Therefore, a resonant state and its time reversal always appear
together for the same $\beta$ and for $k$ and $\overline{k}$,
respectively. 

In addition to the guided modes, Eq.~(\ref{fEq1}) has solutions with
real $\beta$ and real $k$  that
grow exponentially as $z\to \pm \infty$ \cite{Yama,Hanson}. These so-called improper
modes are not physical if they are considered for all $z \in (-\infty,
\infty)$, but on any bounded interval of $z$, they may be realized as
the field profiles of guided modes for some more complicated
waveguides with additional high-index structures at larger values of
$z$. Importantly, the improper modes appear when resonant states reach
their endpoints as $\beta$ is increased \cite{Amgad2}. 

\begin{figure}[htbp]
	\centering
\includegraphics[width=0.8\linewidth]{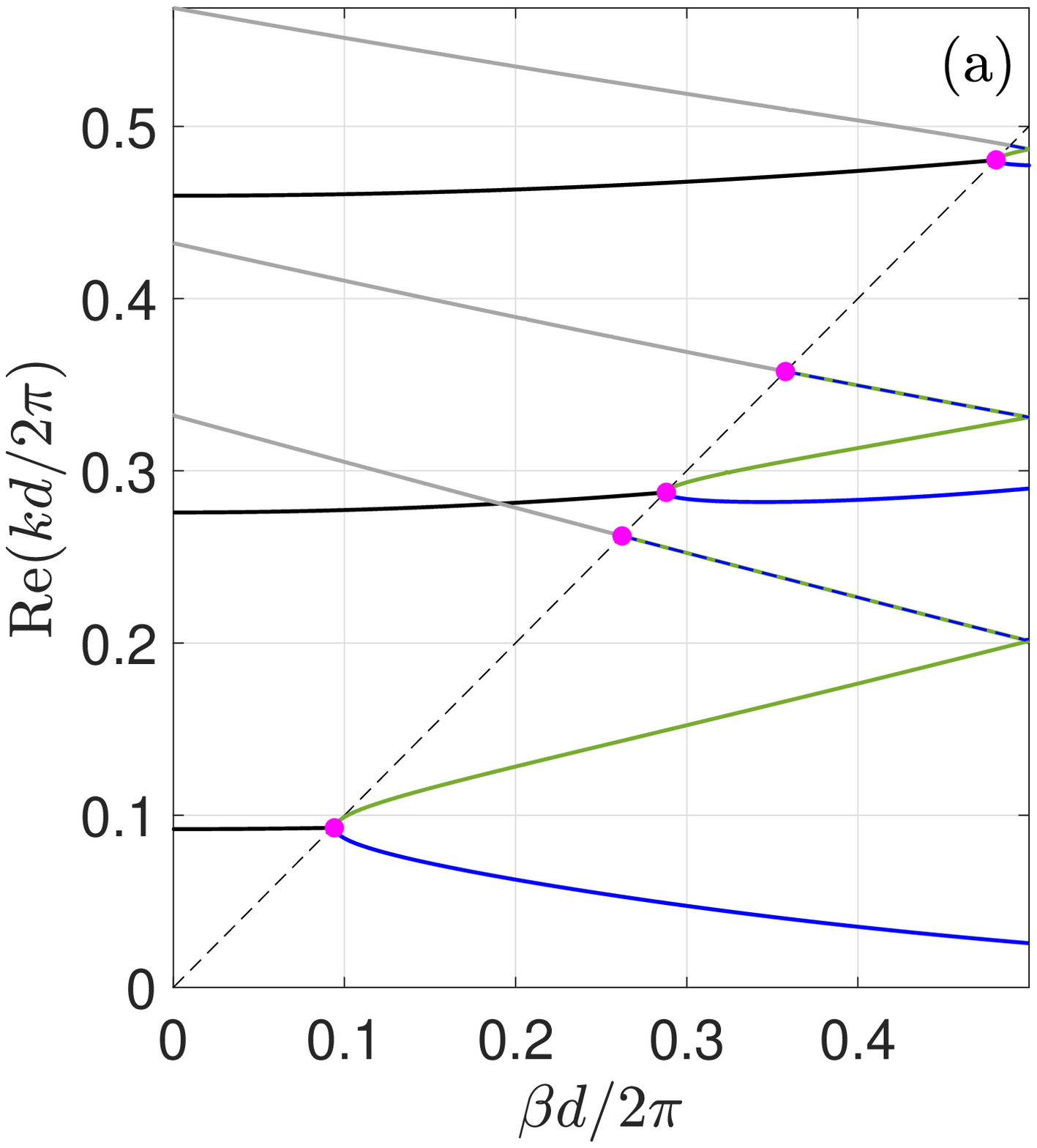}
\includegraphics[width=0.8\linewidth]{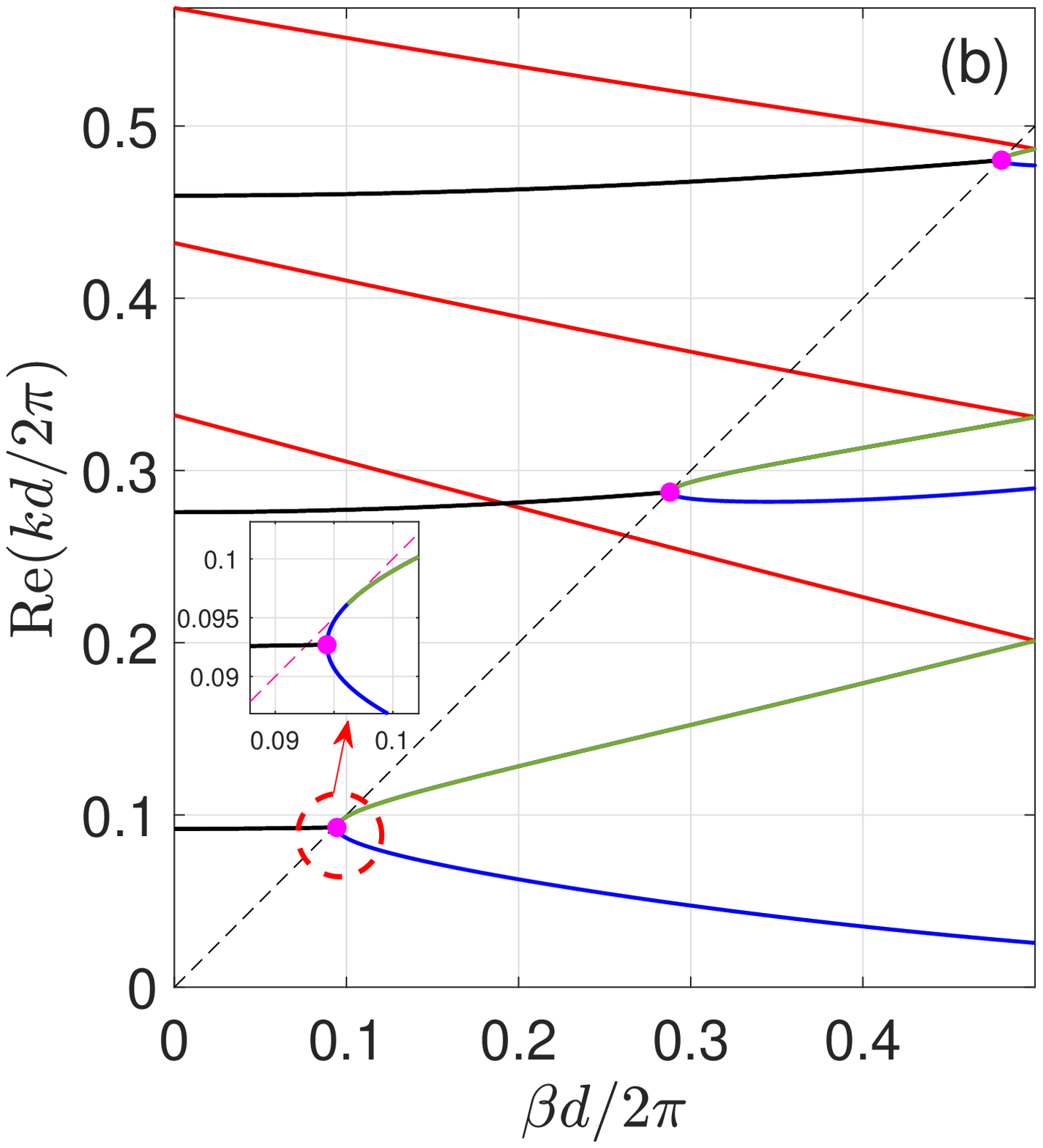}
\caption{Band structures for odd modes of a periodic
  slab (a) and a uniform slab (b). 
 Guided, resonant and improper modes are shown as green, black
  (or gray) and blue curves, respectively. Folded bands of guided
  modes are shown as red curves.  
The purple points are intrinsic EPs.}
\label{fig1cd}
\end{figure}

In Figs.~\ref{fig1cd}(a) and \ref{fig1cd}(b), 
we show the partial band structures of a periodic slab with
$\varepsilon_1 = 11.56$, $\varepsilon_2=11.50$,  
$a = 0.25d$ and $h = 1.6d$, and a uniform
slab with the same $\varepsilon_1$ and $h$, respectively. For both cases,
only a few  odd modes (odd in $z$) are shown, and for resonant states, 
only the real part of $k$ is shown. In Fig.~\ref{fig1cd}(b) for the
uniform slab, 
the dashed line is the light line $k=\beta$, the green curves
represent guided modes that start from points on the light line, the
red curves are folded bands for guided modes with negative propagation
constants given by  $\hat{\beta} = \beta -2\pi/d$, the black curves
represent resonant states (and their time reversals) with endpoints (purple points) below
the light line (visible in the inset), and the blue curves represent
improper modes that diverge as  $z\to \pm  \infty$ \cite{Yama,Hanson,Amgad2}. A
purple point is a branching point between a pair of resonant state and
its time reversal and a pair 
of improper modes. As shown in the inset, the upper branch of the
improper modes converges to the light line  
at the same point where a band of guided modes is formed. 
These purple points can be identified as EPs in a non-Hermitian
EVP formulation consistent with resonant states, their time
reversals, and improper modes. Since they always exist at the
endpoints of dispersion curves for resonant states,  we call them
intrinsic EPs, and denote them as
\[
  {\sf EP}^{(i)}: \quad (R, \overline{R}) \,|\, (I_1, I_2),
\]
where $R$, $\overline{R}$ and $I_1$ and $I_2$ signify a resonant state,
its time reversal, and two improper modes, respectively.
More details can be found in Ref.~\cite{Amgad2}.  

Since $\varepsilon_2$ is close to $\varepsilon_1$, the periodic slab may be regarded as a perturbation of the uniform slab. The solid green, black and blue curves in 
Fig.~\ref{fig1cd}(a) are very close to those in
Fig.~\ref{fig1cd}(b). Thus, the corresponding guided, resonant and
improper modes of the periodic slab are near those of the uniform
slab. On the other hand, a folded band of the uniform slab,
corresponding to a  red curve in Fig.~\ref{fig1cd}(b),  is turned to a
set containing resonant, guided and improper modes shown as solid
gray, dashed green and dashed blue curves in Fig.~\ref{fig1cd}(a).
The dashed green and blue curves nearly coincide and are not
distinguishable in the figure. The set also contains an intrinsic EP
(a  purple point) separating the resonant states and improper modes. It is
below (but very close to) the light line.  Furthermore, a crossing between a black
and a gray curve can be observed in Fig.~\ref{fig1cd}(a). At this
point, the complex frequencies of the two resonant states have the same
real parts, but their imaginary parts are still different. Such a
crossing point can be used as a starting point for searching 
EPs of resonant states. 


\section{A track of exceptional points}
\label{S3}

We are mainly interested in EPs of resonant states  at which two or
more resonant states coalesce.  Typically, a periodic slab with fixed parameters
$\{ \varepsilon_1,  \varepsilon_2, d,  a,  h \}$ does not have such
EPs. A second order EP of resonant states is a codimension-two object corresponding to a point in the plane
of two generic real parameters. Therefore, to find a second order EP,
we need to tune two parameters. Since resonant states form bands that
depend on $\beta$ continuously, 
the Bloch wavenumber $\beta$ can be considered as one parameter.
We choose the thickness $h$ of the slab as the other
parameter. Therefore, if $\varepsilon_1$, $\varepsilon_2$, $d$ and $a$ are fixed,
we expect to find second order  EPs of resonant states as isolated points in the $\beta h$ plane.
In order to find EPs systematically and understand their
dependence on  parameters, we vary  $\varepsilon_2$ continuously and
consider the limit as $\varepsilon_2 \to  \varepsilon_1$. The limit is chosen,
because we expect the conditions for the existence of EPs can be simplified as the
periodic slab approaches a uniform one.

\begin{figure}[htbp]
	\centering 
\includegraphics[width=0.9\linewidth]{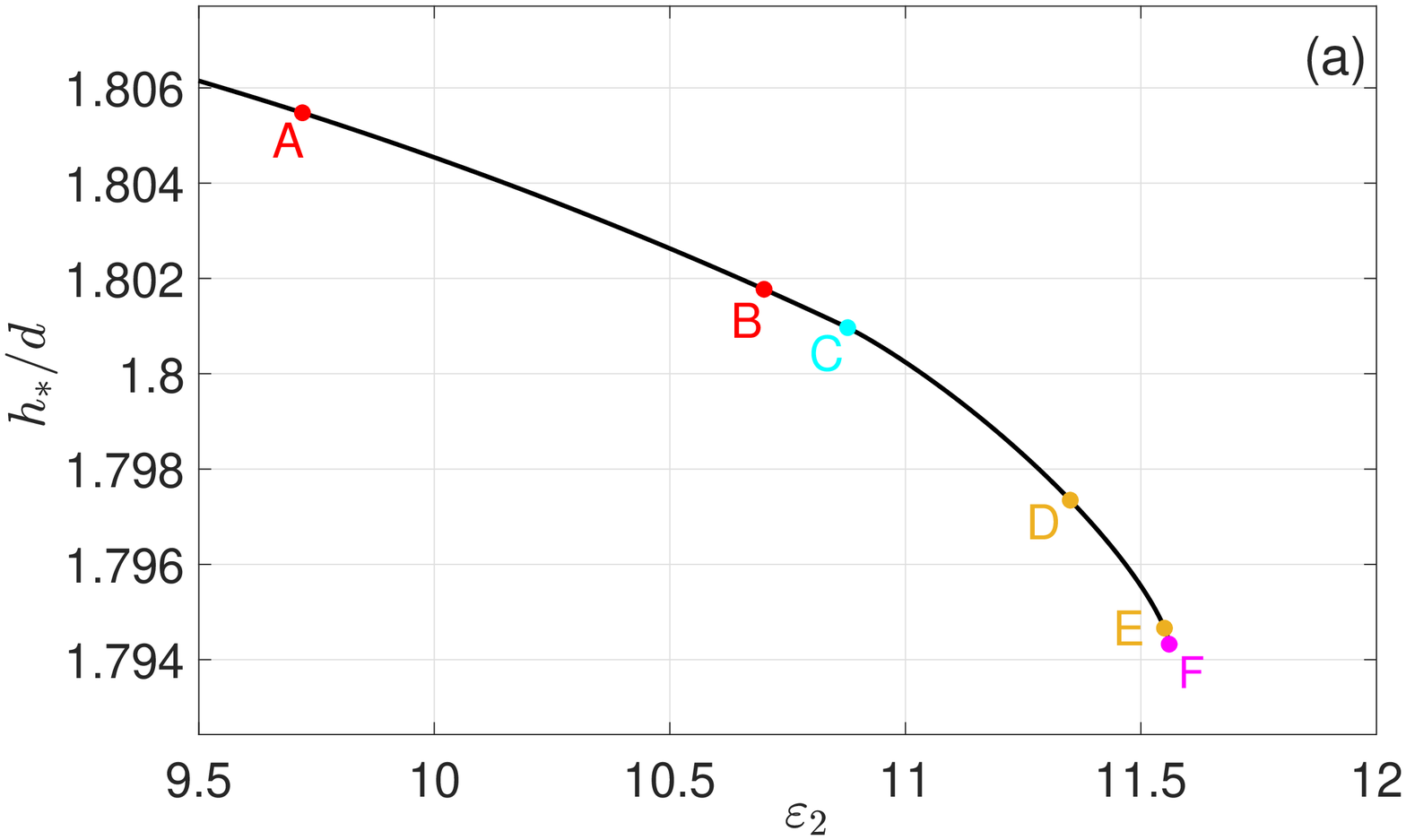}
\includegraphics[width=0.9\linewidth]{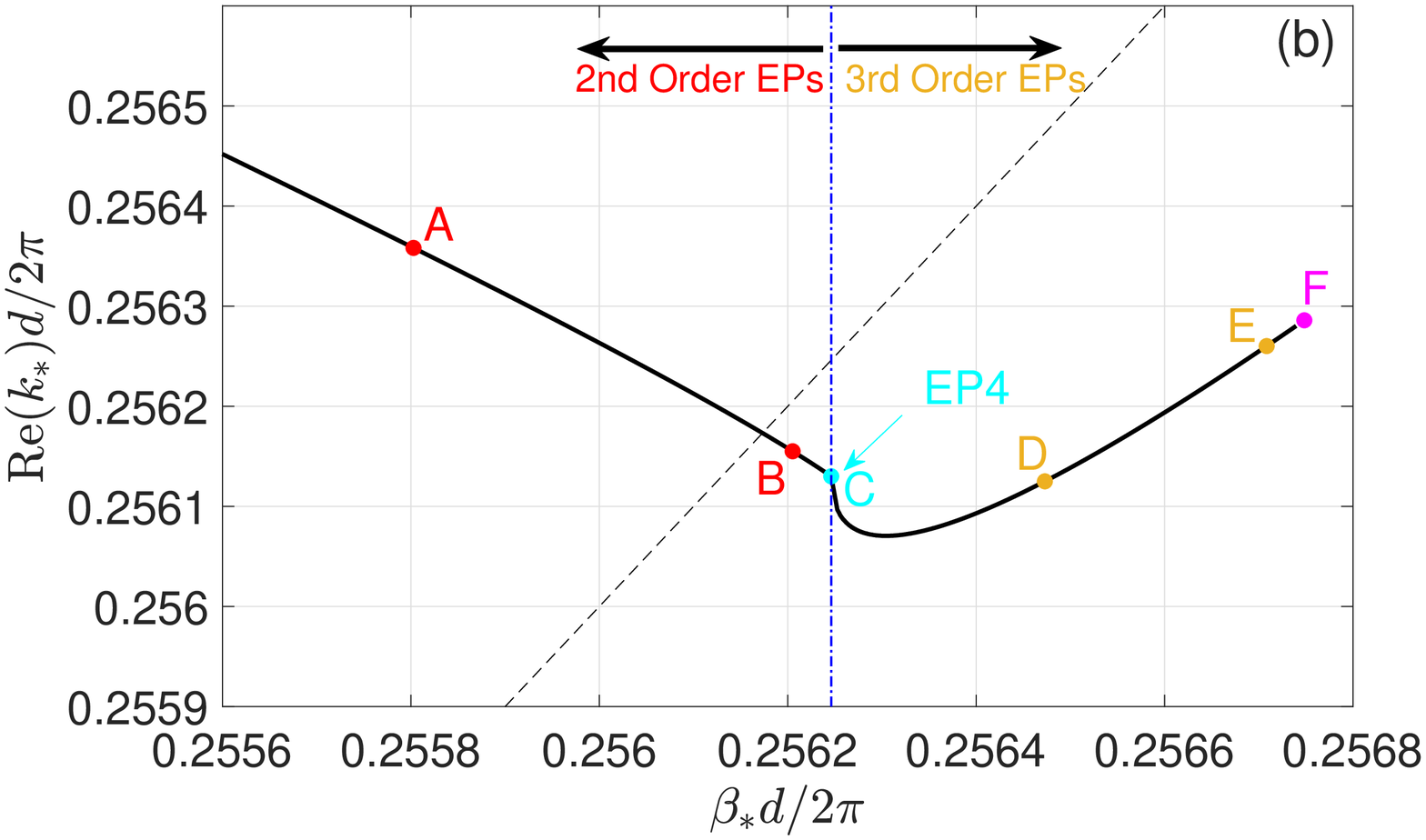}
\includegraphics[width=0.9\linewidth]{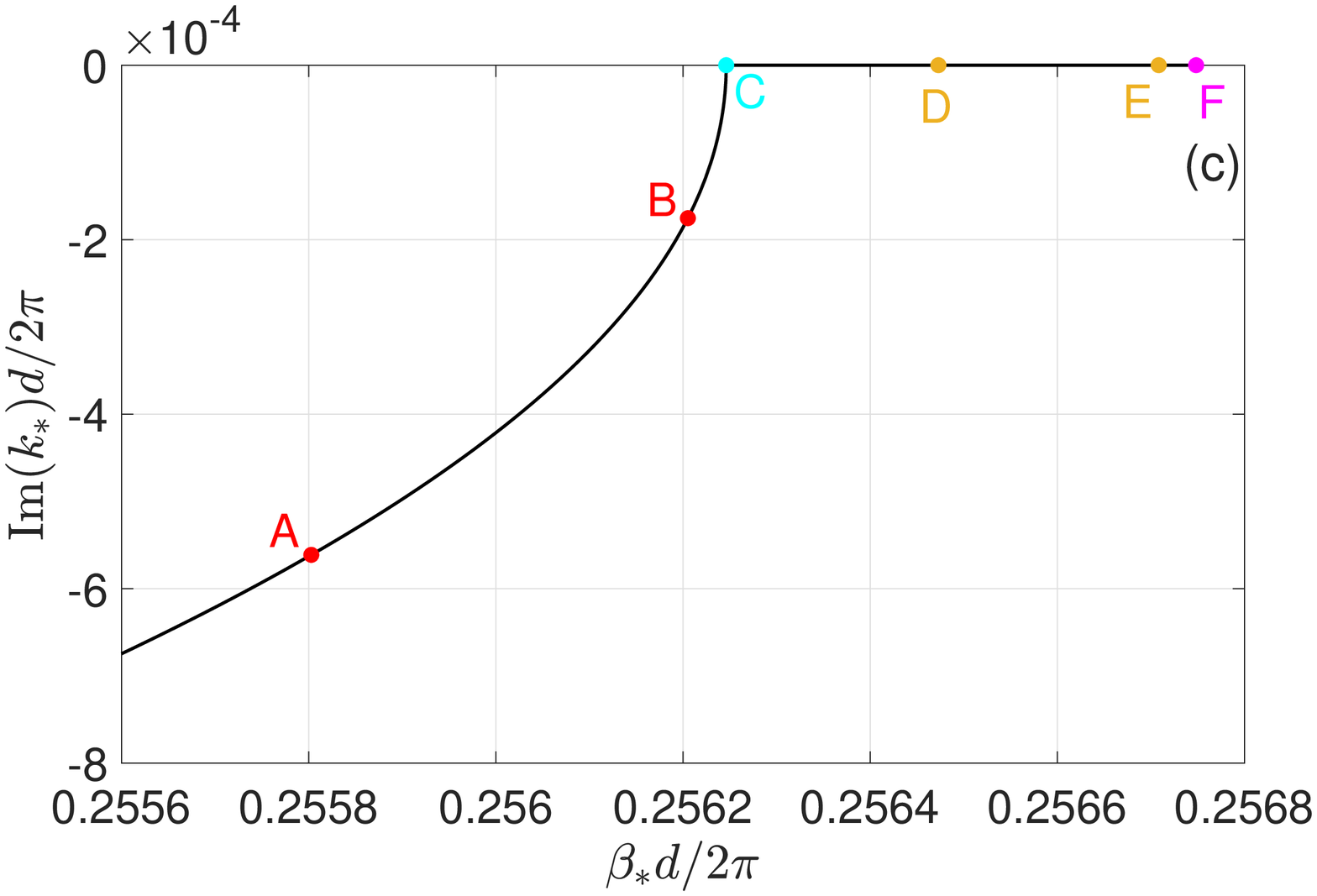}
	\caption{A track of EPs for a periodic slab with an array of
          cylinders. (a) Slab thickness $h_*$ vs. $\varepsilon_2$;
          (b,c) Trajectories of $k_*$ and $\beta_*$ as
          $\varepsilon_2$ varies from 9.5 to $\varepsilon_1=11.56$. 
          ${\sf A}$ and ${\sf B}$ are second order EPs of
          resonant states. ${\sf D}$ and ${\sf E}$ are third
          order EPs with real frequencies. ${\sf C}$ is a fourth
          order EP with a real frequency. ${\sf F}$ is the
          endpoint of the track.}
\label{Fig4.6}
\end{figure}

Using the numerical method described in Appendix, we calculate EPs
for a periodic slab with $\varepsilon_1 = 11.56$,  $a=0.25d$, and a varying
$\varepsilon_2$ from $9.5$ to $\varepsilon_1$. For each $\varepsilon_2$, we determine
the thickness of the slab, denoted as $h_*$, so that structure has an EP,
and denote the Bloch wavenumber and freespace wavenumber 
of the corresponding eigenmode by $\beta_*$ and $k_*$, respectively. 
Multiple solutions exist. We call each continuous family of EPs and associated parameters a track. One particular track is shown in Fig.~\ref{Fig4.6}(a) for $h_*$ as a function of $\varepsilon_2$, and in Figs.~\ref{Fig4.6}(b) and \ref{Fig4.6}(c) for $k_*$ as a complex-valued function of $\beta_*$. 
Of course, both $\beta_*$ and $k_*$ depend on $\varepsilon_2$, but it is
more useful to show $k_*$ as a function of $\beta_*$. It should be
emphasized that the curves in Figs.~\ref{Fig4.6}(b) and
\ref{Fig4.6}(c) are not dispersion curves of a  fixed periodic
structure. Each point on these curves corresponds to a distinct
periodic slab with a unique value of $h_*$.

The track consists of two parts separated by point ${\sf C}$ and ends
at point ${\sf F}$ corresponding to $\varepsilon_2 = \varepsilon_1$. The EPs on the first
part of the track (for smaller values of $\varepsilon_2$, before point ${\sf
  C}$) are second order EPs of resonant states. We denote such an EP by
\[
  {\sf EP2}: \quad (R_1, R_2) \,|\,  (R_1, R_2)
\]
where $R_1$ and $R_2$ signify two resonant states. 
The second part of the track
(between ${\sf C}$ and ${\sf F}$) represents third order EPs at which
three eigenmodes coalesce at a real frequency. The transition point
${\sf C}$ is a fourth order EP with a real $k_*$, and the endpoint
${\sf F}$ is an intrinsic EP satisfying one additional condition. 

\begin{figure}[!htbp]
\centering 
\includegraphics[width=0.9\linewidth]{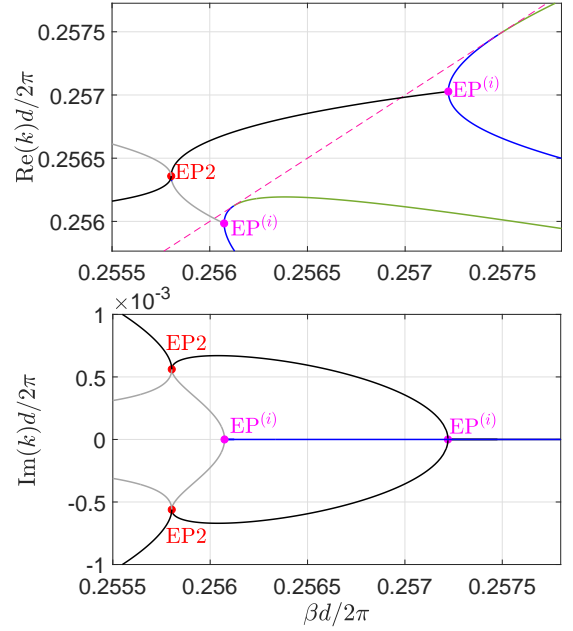}
\caption{Local band structure of a periodic slab with the structural 
  parameters of point ${\sf A}$.
Point ${\sf A}$ is marked as {\sf EP2}.}
\label{FigA}
\end{figure}

On the first part of the track, we choose two points ${\sf A}$ and
${\sf B}$ and show local band structures of the periodic slab at the
associated parameter values. Point ${\sf A}$ corresponds to a periodic
slab with $\varepsilon_2 = 9.72$ and $h_* = 1.805476d$, and a degenerate
resonant state with $\beta_* d=1.607256$ and $k_*d =
1.610746-0.003526i$. 
In Fig.~\ref{FigA},
we show some dispersion curves near this degenerate resonant state
marked as ${\sf EP2}$. The real and imaginary parts of $k$ are shown
in the upper and lower panels, respectively.  It is clear that two
resonant states, shown as black and gray curves, collapse at the EP
with standard square-root splittings for both the real and imaginary
parts of $k$ and on both sides of $\beta_*$. For $\beta > \beta_*$,
the resonant states exist continuously until they reach their ends
below the light line at the intrinsic EPs, marked as ${\sf EP^{(i)}}$
in the figure. The lower panel of Fig.~\ref{FigA} show imaginary parts
of both $k$ and $\overline{k}$ corresponding to the resonant states
and their time reversals. The EP in the upper half plane, also marked
as {\sf EP2}, is the one for the time reversals. 

\begin{figure}[!htbp]
	\centering 
	\includegraphics[width=0.9\linewidth]{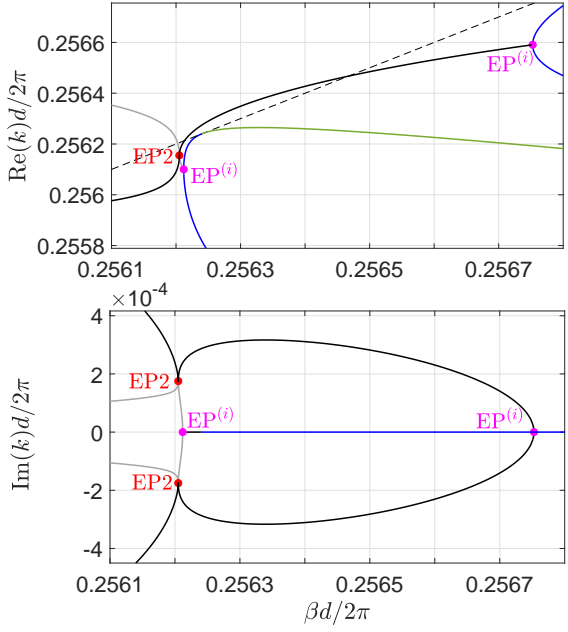}
	\caption{Local band structure of a periodic slab with the structural 
  parameters of point ${\sf B}$. 
Point ${\sf B}$ is marked as {\sf EP2}.}
 \label{FigB}	
\end{figure}

Point ${\sf B}$ corresponds to a periodic slab with $\varepsilon_2 = 10.70$
and $h_* =  1.801773d$, and a degenerate resonant state with $\beta_*d
= 1.609784$ and $k_*d = 1.609470-0.001102i$. 
The local band structure for this periodic slab is shown in Fig.~\ref{FigB}.
The resonant state for this EP (marked as {\sf EP2} in Fig.~\ref{FigB}) 
is below the light line, and is very close to one intrinsic EP. In the
lower panel of Fig.~\ref{FigB}, both $k$ and $\overline{k}$ are
shown. It can be seen that the EP of the resonant states and the EP of 
the time reversals are very close to each other. 
  
\begin{figure}[!htbp]
\centering
\includegraphics[width=1.0\linewidth]{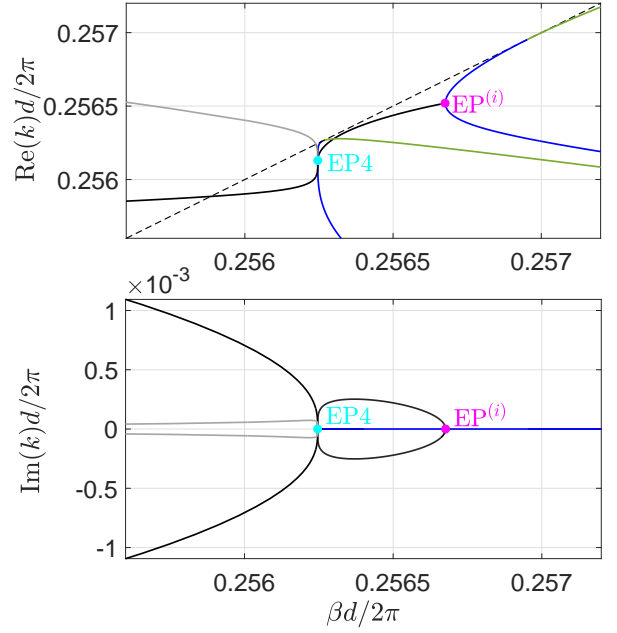}
\caption{Local band structure of a periodic slab with the structural 
  parameters of point ${\sf C}$. 
  Point ${\sf C}$ is marked as {\sf EP4}.}
\label{FigC} 	
\end{figure}

Point ${\sf C}$ is a fourth order EP obtained when two second order 
EPs (for the resonant states and their time reversals, respectively)
collapse with the nearby intrinsic EP.  
It corresponds to a periodic slab
with $\varepsilon_2 \approx 10.877538$ and $h_*= 1.8009676d$, and a 
degenerate eigenmode with $\beta_*d = 1.6100415$d and $k_*d=
1.6093126$. It is a fourth order EP, since four modes coalesce at this
point. The four modes are two resonant states and their 
time reversals for $\beta < \beta_*$,  and 
one resonant state, its time reversal, and two improper modes 
for $\beta > \beta_*$. We denote this fourth order EP as
\[
  {\sf EP4}: \quad (R_1, R_2, \overline{R}_1, \overline{R}_2) \, | \,
  (R, \overline{R}, I_1, I_2)
\]
In Fig.~\ref{FigC}, we show the band structure for the periodic slab
with $\varepsilon_2$ and $h_*$ given above. 
The resonant states and their time reversals are shown as solid black or
gray curves, and the improper modes are shown as solid blue curves.
For $\beta > \beta_*$, there is still one resonant mode and its time
reversal, and they exist until they reach an intrinsic EP. Moreover, the upper
branch of improper modes approaches the light line and ends at the
same point where a branch of guided modes emerges. 

\begin{figure}[!htbp]
 \centering 
 \includegraphics[width=\linewidth]{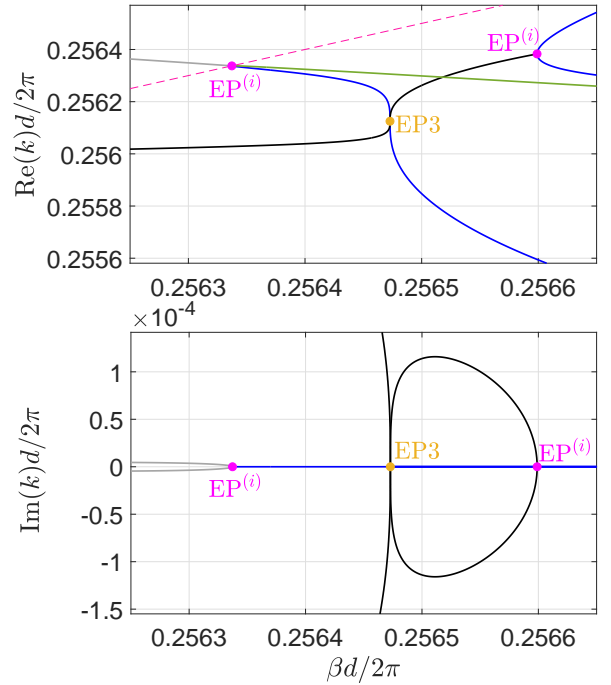}
\caption{Local band structure of a periodic slab with the structural 
  parameters of point ${\sf D}$. 
  Point ${\sf D}$ is marked as {\sf EP3}.}
\label{FigD}	 
\end{figure}

As $\varepsilon_2$ is further increased, an intrinsic  EP appears near the light line. 
As usual, it connects a resonant state and its time reversal (for
smaller values of $\beta$) with two improper modes (for larger values of
$\beta$). One improper mode can 
collapse with another  resonant state and its time reversal to form a
third order EP. It turns out that as $\beta$ passes through its value
at this third order EP, there are still a resonant state, its time reversal and
an improper mode. Therefore, we denote this third order EP by 
\[
  {\sf EP3}: \quad (R, \overline{R}, I) \,|\, (R, \overline{R}, I).
\]
As an example, we consider point ${\sf D}$ which corresponds to
a  periodic slab with $\varepsilon_2 = 11.35$ and $h_* = 1.7973466d$, and a
triply degenerate eigenmode with $\beta_*d = 1.6114666$ and $k_*d= 1.6092809$. 
The local band structure of the periodic slab is shown in Fig.~\ref{FigD}.
We observe that this third order EP is connected with an intrinsic EP 
(at a smaller value of $\beta$)  by a branch of improper modes, and is
connected to another intrinsic EP (at a larger value of $\beta$) by a
branch of resonant states (and their time reversals). 

\begin{figure}[!htbp]
  \centering 
  \includegraphics[width=\linewidth]{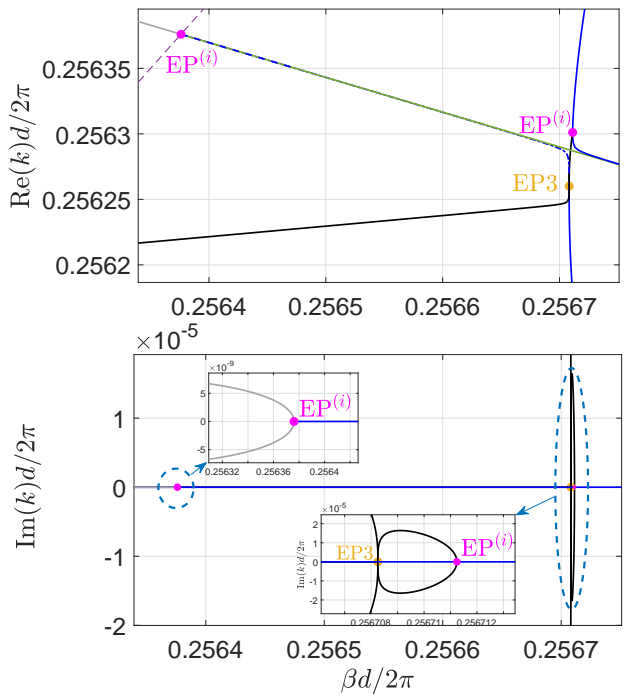}
  \caption{Local band structure of a periodic slab with the structural 
  parameters of point ${\sf E}$. 
  Point ${\sf E}$ is marked as {\sf EP3}.}
\label{FigE}	
\end{figure}

Point ${\sf E}$ is another third order EP corresponding to a periodic slab with
$\varepsilon_2 =11.55$ and $h_* = 1.7946619d$, and a degenerate eigenmode
with $\beta_*d = 1.6129455$ and $k_*d = 1.61013$. The local
band structure of this periodic slab is shown in Fig.~\ref{FigE}.
We can see that the third order EP is very close to an
intrinsic EP that exists at a slightly larger value of $\beta$. There
is a  very short branch of resonant states connecting the third
order EP with this intrinsic EP. Another intrinsic EP can be found
near the light line. It is connected with the third order EP by a
branch of improper modes that nearly coincide with a branch of guided
modes. 

\begin{figure}[!htbp]
  \centering 
  \includegraphics[width=\linewidth]{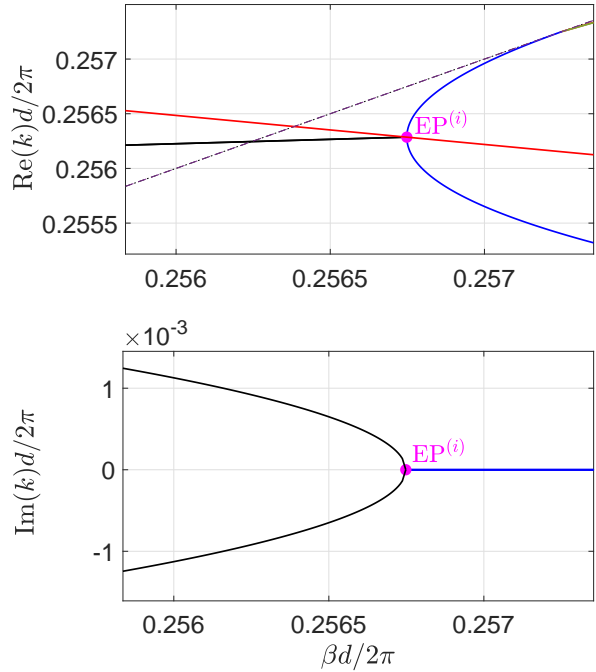}
  \caption{Local band structure of a uniform slab with the structural
    parameters of point ${\sf F}$. 
    A folded band of 
    guided modes is shown as the red curve. Point ${\sf F}$ coincides
    with the intrinsic EP marked as ${\sf EP}^{(i)}$.}
  \label{FigF}	
\end{figure}

The final point ${\sf F}$ is obtained when the third order EP and the
intrinsic EP at a slightly larger value of $\beta$ shown in
Fig.~\ref{FigE}, merge together as $\varepsilon_2 \to \varepsilon_1$. It corresponds to a uniform slab
with $h_* = 1.7943252d$ and a  degenerate eigenmode with 
$\beta_*d =1.613197$ and $ k_*d=  1.6102917$. As $\varepsilon_2$ tends to
$\varepsilon_1$, the branch of improper modes connecting the intrinsic EP
near the light line and the third order EP as shown in Fig.~\ref{FigE},
collapses with a branch of guided modes, a new intrinsic EP is formed
connecting a branch of resonant states and their time reversals on one side  with
the two remaining branches of improper modes on the other side. As
shown in Fig.~\ref{FigF},  
point ${\sf F}$ corresponds to a uniform slab where a folded band of
guided modes intersects with an intrinsic EP.

\section{Limiting points and additional tracks}
\label{S4}

Numerical results of the previous section suggest that EPs follow a track
with a limiting point corresponding to a uniform slab with a folded band
of guided modes intersecting an intrinsic EP. In this section, we 
analyze uniform slabs satisfying this condition and verify that they indeed
give rise to tracks of EPs.

For a uniform slab 
with dielectric constant
$\varepsilon_1$ and thickness $h$, a guided mode is 
$ u(y,z) = \hat{\phi}(z) \exp ( i \hat{\beta} y) $, where $\hat{\beta}$ is
the propagation constant and $\hat{\phi}$ tends to zero as $z \to \pm
\infty$.  With a fictitious period $d$ in the $y$ direction, the mode 
can be written as 
\[
  u(y,z) =  \phi(y,z) e^{i \beta y}, 
  \quad \phi(y,z)  = \hat{\phi}(z)  e^{-i 2\pi p y/d}
  \]
for an integer $p$, such that   $\beta = \hat{\beta} +  2\pi p/d \in
[-\pi/d, \pi/d]$. A folded band corresponds to a set of modes with
a nonzero $p$ and $\beta \in [-\pi/d, \pi/d]$.
If $p=1$,  then $\hat{\beta} = \beta - 2\pi/d \in
[-3\pi/d, -\pi/d]$.

If the guided mode is odd in $z$, then 
\[
  \hat{\phi}(z) = \begin{cases}
    -\exp[  \hat{\gamma}_0 (z + h/2) ], & z <  -h/2 \cr 
    \sin (\hat{\gamma}_1 z) / \sin(\hat{\gamma}_1 h/2), &  |z| < h/2, \cr 
    \exp [  -\hat{\gamma}_0 (z-h/2)], & z > h/2, 
    \end{cases}
  \]
where $\hat{\gamma}_1 = (k^2 \varepsilon_1 - \hat{\beta}^2)^{1/2}$ and 
$\hat{\gamma}_0 = (\hat{\beta}^2 - k^2)^{1/2}$, and the dispersion
curves can be determined from
\begin{equation}
  \label{slabodd}
  \cot  \left( \hat{\gamma}_1 h / 2 \right) + \hat{\gamma}_0 / \hat{\gamma}_1   = 0. 
\end{equation}
The odd guided modes form bands with starting points on the light line
given by 
\[
  k=|\hat{\beta}| = \frac{ (2n-1)\pi}{ h \sqrt{ \varepsilon_1 -1}}, \quad n=1,
  2, \cdots
\]
The $n$-th band can be determined from
\begin{equation}
  \label{oddn}
  \hat{\gamma}_1 h / 2  + \mbox{arccot} \left(  \hat{\gamma}_0 /
    \hat{\gamma}_1  \right)   = n \pi. 
\end{equation}

An improper mode of the uniform slab is similarly given by  $u(y,z) = \phi(z) e^{ i
  \beta y}$, but $\phi$ grows  exponentially as $z \to
\pm \infty$. If the mode is odd in $z$, then
\[
  \phi(z) = \begin{cases}
    -\exp[  -\gamma_0 (z + h/2) ], & z <  -h/2 \cr 
    \sin ( \gamma_1 z) / \sin(\gamma_1 h/2), &  |z| < h/2, \cr 
    \exp [  \gamma_0 (z-h/2)], & z > h/2, 
  \end{cases}
\]
where $\gamma_1 = (k^2 \varepsilon_1 - \beta^2)^{1/2}$ and 
$\gamma_0 = (\beta^2 - k^2)^{1/2}$. The dispersion relations can be
obtained by solving 
\begin{equation}
  \label{impodd}
  \cot  \left(  \gamma_1 h / 2 \right) - \gamma_0 /
    \gamma_1 = 0.
  \end{equation}
The $m$-th band of improper modes emerges from the same point on the
light line as the $m$-th guided mode, and it can be determined from
\begin{equation}
  \label{oddm}
  \gamma_1 h / 2  - \mbox{arccot} \left( \gamma_0 /
    \gamma_1  \right)   = (m-1) \pi
\end{equation}
for $m\ge 1$. On each band, $\beta$ has a minimum which 
corresponds to an intrinsic EP. Taking the derivative with respect to $k$ for 
Eq.~(\ref{impodd}) and setting $d\beta/dk=0$, we obtain the following
condition for  intrinsic EPs:
\begin{equation}
  \label{oddmin}
   \beta^2 = h \varepsilon_1 k^2 \gamma_0 /2.
\end{equation}

Therefore, if a folded band of guided modes intersects with an intrinsic EP, we can
find $h$, $k$ and $\beta$ by
solving Eqs.~(\ref{oddn}), (\ref{oddm}) and (\ref{oddmin}). The band
indices must satisfy $m > n \ge 1$. 

The case for even modes (even in $z$) is similar. The $n$-th guided
mode with propagation constant $\hat{\beta}$ and the $m$-th improper
mode with propagation constant $\beta$ satisfy 
\begin{eqnarray}
  \label{egun}
&& \hat{\gamma}_1  h/ 2  -\arctan \left ( 
   \hat{\gamma}_0 /  \hat{\gamma}_1 \right) = n \pi, \\
  \label{eimm}
&& \gamma_1 h / 2 +\arctan  \left( \gamma_0/ \gamma_1
   \right) = m\pi
\end{eqnarray}
respectively, where $n \ge 0$ and $m \ge 1$. 
An intrinsic EP corresponds to a minimum of $\beta$ as a function of
$k$ for a band of improper modes. It turns out that the condition 
$d\beta/dk=0$ for even improper modes gives the same
Eq.~(\ref{oddmin}).
Therefore, if a folded band of even guided modes intersects an
intrinsic EP,  Eqs.~(\ref{oddmin}), (\ref{egun}) and (\ref{eimm}) are
satisfied. 

\begin{figure}[htbp]
\centering 
\includegraphics[width=\linewidth]{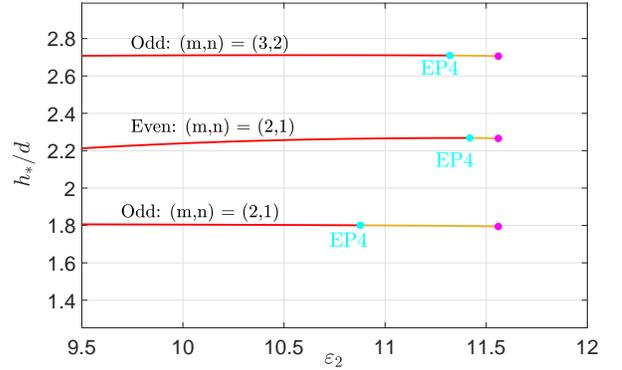}
\caption{Three different tracks of EPs for a periodic slab with an
  array of circular cylinders. The slab thickness $h_*$ is shown as functions
  of $\varepsilon_2$.}
\label{Fig4.14}	
\end{figure}

In Table~\ref{Tab41},
\begin{table}[htbp]
\centering 
\caption{Endpoints for tracks of EPs: uniform slabs with a folded band
  intersecting an intrinsic EP.}
\label{Tab41} 
\begin{tabular}{ccccc}\hline\hline 
		$(m,n)$&$h_*/d$ & $\beta_*d/2\pi $&$k_*d/2\pi$ & parity\\\hline 
		$(2,1)$&1.794325& 0.256748 &  0.256286& odd\\
		$(3,1)$&3.226579& 0.238265 &  0.238112 & odd\\
          $(3,2)$&2.705984& 0.284104 &  0.283922& odd\\
          $(1,0)$&1.273128  &  0.240640  & 0.239650 & even \\
          $(2,0)$&      2.658020& 0.231293 &  0.231060& even \\
		$(2,1)$&2.265665& 0.271347 &  0.271074& even\\
		$(3,1)$&3.752632& 0.245890 &  0.245781&even\\
		$(3,2)$&3.124956& 0.295280 &  0.295148&even\\ 
		\hline\hline 
\end{tabular}
\end{table}
we list a few solutions (denoted as $h_*$, $\beta_*$ and $k_*$) of
Eqs.~(\ref{oddn}), (\ref{oddm}) and (\ref{oddmin}) for odd modes and 
Eqs.~(\ref{oddmin}), (\ref{egun}) and (\ref{eimm}) for even modes,
assuming the folded bands are defined by $p=1$. It is
clear that the final point ${\sf F}$ of previous section corresponds to the case
$(m,n)=(2,1)$ for odd modes.
The other solutions also correspond to tracks of EPs. Therefore, we
can assign an integer pair $(m,n)$ and parity (odd
or even) to each track. In Fig.~\ref{Fig4.14},
we show slab thickness $h_*$ as functions of $\varepsilon_2$ for three
tracks of EPs. The curve marked with ``Odd: (m,n)=(2,1)'' is identical
to that in Fig.~\ref{Fig4.6}(a). All three tracks have the same
property. That is, the first part corresponds to second order EPs of
resonant states, the second part corresponds to third order EPs, and
the transition point is a fourth order EP. 

In Ref.~\cite{Amgad1}, we analyzed EPs for a periodic slab
with two rectangular segments in each period, and followed EPs as the
width of one segment is decreased. Due to numerical difficulties, we
only calculated second order EPs of resonant states up to the light
line, and obtained an approximate condition for the limiting uniform
slab, namely, a folded band of guided modes intersecting the starting
point of guided modes on the light line. For the case of odd modes
and $(m,n)=(2,1)$, the approximate condition gives a slab thickness
$h_*^{\rm app} = 1.801753$, and the relative error
$|h_*^{\rm app} - h_*|/h_*$ is about $0.4\%$.

\section{Conclusion}
\label{S5}

Due to the many interesting properties and potential applications, EPs
for photonic structures are being intensively studied by many
researchers. Periodic structures have a special advantage related to
the degree of freedoms  provided by the Bloch wavevector,
so that a typical second order EP of resonant states can be found
without tuning structural parameters (for bi-periodic structures) or
tuning only one structural parameter (for structures with a single
periodic direction). In this paper, we analyzed EPs for a dielectric
slab containing  a periodic array of cylinders. As in our previous work
\cite{Amgad1}, we aim to develop a clear picture about EPs in
parameter space  by analyzing  the limit when the periodic
structure approaches a uniform slab. It appears that EPs stay on
tracks that can be indexed by a pair of integers and the even/odd
parity. The EPs on each track exhibit an interesting transition from
second order EPs of resonant states (with complex frequencies) to a
special kind of third order EPs with real frequencies, via a fourth
order EP also with a real frequency. We also found a precise condition
for uniform slabs that correspond to the endpoints of tracks.

Our approach is applicable to more general structures. Instead of
calculating isolated EPs by searching parameters locally, we can 
follow the EPs continuously by tuning parameters towards a simper
structure (such as a uniform slab) and then analyze the simpler
structure analytically. For a general 2D periodic structure with a reflection
symmetry in $z$, we expect to obtain the same condition for the
limiting uniform slab.  If the periodic structure does not have the
reflection symmetry in $z$, the modes can no longer to separated as
even and odd ones,  a slightly different condition can be expected.
Moreover, a similar study can be carried out for  bi-periodic structures 
such as photonic crystal slabs. 
Our work improves the theoretical understanding on EPs and can provide
useful guidance for their practical applications.

\begin{acknowledgments}
  The authors acknowledge support from the Research Grants
Council of Hong Kong Special Administrative Region, China (Grant
No. CityU 11305518).
\end{acknowledgments}

\section*{Appendix}

To find different kinds of Bloch modes satisfying Eq.~(\ref{fEq1}), we can formulate a linear EVP (for
eigenvalue $k$ or $k^2$)  on one period of the structure given by
$|y| < d/2$, but it is necessary to truncate $z$ by some 
technique such as the perfectly matched layer (PML) and 
discretize the truncated 2D domain. This leads to a linear matrix EVP, but the size of the matrix is relatively large.
Alternatively, we can formulate a nonlinear EVP
 \begin{equation}
   \label{fEq4}
    A(\beta,k) w = 0
 \end{equation}
 where the eigenvalue $k$ appears in $A$ implicitly. The advantage of
 the  nonlinear EVP formulation is that $A$ can be approximated by a
 much smaller matrix and there is no need to truncate $z$. We use a nonlinear
 EVP formulation based on a cylindrical wave expansion in the square
 given by $|y|<d/2$ and $|z|< d/2$, a plane wave expansion in rectangles
 given by $|y| < d/2$ and $ d/2 < |z| < h/2$, and another plane wave
 expansion in the infinite domains given by  $|y| < d/2$ and $|z| >
 h/2$. For the cylindrical wave expansion, we need to impose the
 quasi-periodic condition, since $u$ is a Bloch mode given by
 Eq.~(\ref{fEq2}). The plane wave expansions satisfy the quasi-periodic
 condition automatically. The system (\ref{fEq4}) is obtained when the
 expansions are properly truncated, and continuity conditions are
 imposed at $z=\pm d/2$ and $z=\pm h/2$. If we consider even and odd
 modes separately, the size of matrix $A$ can be  further reduced.  To
 find the dispersion curves, such as those in Fig.~\ref{fig1cd}(a),
 we solve $k$ for given $\beta$ from the condition that $A$ is a
 singular matrix. Typically, we use the condition $\lambda_1 (A) = 0$
 where $\lambda_1$ is the linear eigenvalue of $A$ with the smallest magnitude. 

To find a second order EP of resonant states, we realize that $A$
depends on structural parameters such as the slab thickness $h$, and
solve $h$, $\beta$ and $k$ from
\[
  \lambda_1 = \frac{d \lambda_1}{dk} = 0.
\]
A theoretical justification for the second condition $d\lambda_1/dk=0$ is given in
Ref.~\cite{Amgad3}. 
Since $\lambda_1$ is in general complex, the above two conditions
are equivalent to four real conditions. The unknowns $h$ and $\beta$
are real and $k$ is complex, and thus they are equivalent to four real
unknowns. A third order EP in the second part of each track always has 
a real frequency. We solve real $h$, $\beta$ and $k$ from the
following three conditions 
\[
  \lambda_1 = \frac{d \lambda_1}{dk} = \frac{d^2 \lambda_1 }{dk^2} =0.
\]
It turns out that in the regime where such third order EPs exist,
$\lambda_1$ is always real.

\bibliography{MyBib}

\end{document}